\documentclass[journal = jpclcd, manuscript = letter,layout=twocolumn]{achemso}

\usepackage{amssymb,amsfonts,amsmath}

\usepackage{color}
\usepackage{ulem}

\title{Femtosecond Intersystem Crossing in the DNA Nucleobase Cytosine}

\author{Martin Richter}
\affiliation{Institute of Physical Chemistry, Friedrich Schiller
University Jena, Helmholtzweg 4, 07743 Jena, Germany}
\author{Philipp Marquetand}
\affiliation{Institute of Theoretical Chemistry, University of
Vienna, W\"ahringer Str. 17, 1090 Vienna, Austria}
\author{Jes\'us Gonz\'alez-Vazquez}
\author{Ignacio Sola}
\affiliation{Departamento de Qu\'imica F\'isica I,
Universidad Complutense, 28040 Madrid, Spain}
\author{Leticia Gonz\'alez}%
\email{leticia.gonzalez@univie.ac.at}
\affiliation{Institute of Theoretical Chemistry, University of
Vienna, W\"ahringer Str. 17, 1090 Vienna, Austria}

\begin{document}

\makeatletter
\setlength\acs@tocentry@height{5.1cm}
\setlength\acs@tocentry@width{5.1cm}
\makeatother
\begin{tocentry}
\includegraphics[width=5.1cm]{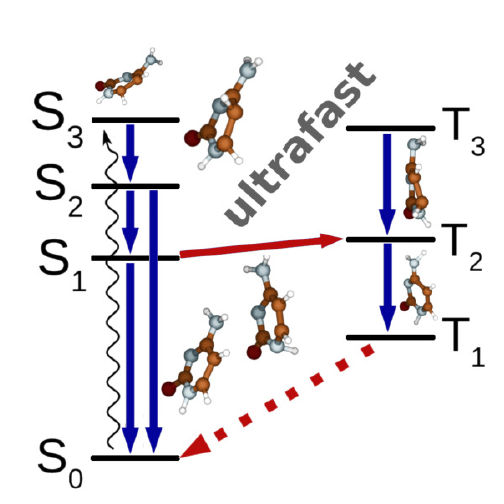}
\end{tocentry}

\maketitle


\begin{abstract}
Ab initio molecular dynamics including non-adiabatic and spin-orbit couplings on equal footing is used to unravel the deactivation of cytosine after UV light absorption. Intersystem crossing (ISC) is found to compete directly with internal conversion in tens of femtoseconds, thus making cytosine the organic
compound with the fastest triplet population calculated so far. It is found that close degeneracy between singlet and triplet states can more than compensate for very small spin-orbit couplings, leading to efficient ISC. The femtosecond nature of the intersystem crossing process highlights its
importance in photochemistry and challenges the conventional view that large singlet-triplet couplings are required for an efficient population flow into triplet states. These findings are important to understand DNA photostability and the photochemistry and dynamics of organic molecules in general.
\end{abstract}

\newpage

The interaction of DNA and RNA with radiation, from mobile-phone emissions~\cite{Nylund2006P} to UV wavelengths~\cite{Crespo-Hernandez2005NAT}, has enthralled the scientific community for years due to its implications in photodamage~\cite{Shukla2008}. Of particular interest is to understand photostability, i.e. the relaxation mechanisms that bring DNA~\cite{Schultz2004SCI,Abo-Riziq2005PNAS,Barbatti2010PNAS} to the ground state before any other photoreaction can occur. This means that, instead of fluorescence or phosphorescence, the electronic energy provided upon photoexcitation in DNA is transferred to the nuclear degrees of freedom of the molecular system in different ways. It is precisely the atomistic description of these different relaxation pathways that is still heavily discussed in the literature.
In the last years it has been clearly established that excited states of isolated DNA nucleobases undergo ultrafast internal conversion (IC) allowing for an efficient radiationless decay towards lower-lying electronic states~\cite{Peon2001CPL,Pecourt2000JACS,Kang2002JACS,Crespo-Hernandez2004CR,Ullrich2004PCCP,
Canuel2005JCP,Kwok2006JACS,Hare2007PNAS,Vaya2010JACS}.
The role of intersystem crossing (ISC) in the process of photostability is, however, much less discussed~\cite{Hare2006JPCB,Hare2007PNAS,Hare2008CP,Kwok2008JACS,Etinski2009JPCA}, probably because it is thought to be a much slower process in comparison to IC~\cite{McQuarrie1997} and also because the quantum yields of triplet states population in DNA and RNA nucleobases are generally very small and thus difficult to access from the experimental point of view.~\cite{Crespo-Hernandez2004CR,Hare2007PNAS,Cadet1990BP}
We note, however, that ultrafast time scales for ISC in other organic molecules
have been experimentally reported or predicted before.~\cite{Cavaleri1996CPL,Tamai1992CPL,Aloise2008JPCA,Crespo-Hernandez2008JPCA,Zugazagoitia2009JPCA,
Parker2009CPL,Reichardt2009JCP,Reichardt2010JPCL,Reichardt2010CC,Reichardt2011JPCB,
Minns2010PCCP,Penfold2010CP,Etinski2011JCP,Ghosh2012JPCA,Yang2012JCP}

In this work we present the first excited state dynamical study of a DNA nucleobase including singlet and triplet states. Such simulations are done using the newly developed surface-hopping method
\texttt{SHARC}~\cite{Richter2011JCTC}. \texttt{SHARC} stands for
\textsc{s}urface \textsc{h}opping including \textsc{ar}bitrary \textsc{c}ouplings. With \texttt{SHARC} one can treat non-adiabatic and spin-orbit couplings (which mediate IC and ISC, respectively) on equal footing. The applicability of \texttt{SHARC} to include spin-orbit as well as dipole couplings is
documented in Refs.~\cite{Richter2011JCTC,Marquetand2011FD,Bajo2011JPCA}. Here, \texttt{SHARC} is employed to investigate the role of the triplet states in the deactivation of cytosine within the framework of nonadiabatic molecular dynamics based on ab initio multiconfigurational methods. Such study is necessary to provide a mechanistic insight that goes beyond what can be learned from quantum chemical calculations alone.

Cytosine presents three tautomers: enol-, keto- and keto-imine-cytosine.
Keto-cytosine is the biological relevant tautomer found in the DNA's nucleotides linked to the deoxyribose sugar moiety and the only one for which a crystalline structure exists~\cite{Barker1964AC}.
Therefore, here we focus on keto-cytosine. Several stationary
points~\cite{Shukla2008,Blancafort2008,Matsika2011FD} of keto-cytosine, including
two-~\cite{Ismail2002JACS,Merchan2003JACS,Tomic2005JPCA,Blancafort2004JACS,
Blancafort2007PP,Kotur2011JCP} and three-state~\cite{Blancafort2004JPCA,Kistler2007JPCA,Kistler2008JCP} conical intersections involved in the process of IC have been calculated with ab initio methods. Time-dependent calculations have indicated that the dynamical behavior of cytosine after photoexcitation is one of the most complicated among nucleobases, involving delocalization of the excited wave packet and relaxation through multiple competing pathways in the singlet excited state manifold~\cite{Barbatti2010PNAS,Hudock2008CPC,Gonzalez-Vazquez2010CPC,Barbatti2011PCCP}. The possible triplet state formation via ISC along the internal conversion pathway
of excited singlet keto-cytosine has been discussed by Merch\'{a}n et
al.~\cite{Merchan2005JACS,Gonzalez-Luque2010JCTC} in the light of quantum
chemical calculations.

Our ab initio molecular dynamic simulations are performed on seven states simultaneously: four singlets and three triplets. Energies, energy gradients, non-adiabatic and spin-orbit couplings are computed on-the-fly using the state-average Complete Active Space Self-Consistent Field (CASSCF) method~\cite{Werner1985JCP,Knowles1985CPL}. Further details are given in the Supporting Information (SI). The first singlet excited state, S$_1$, has $\pi \pi^*$ character at the equilibrium geometry and it is bright while the states higher in energy, S$_2$ and S$_3$, correspond to dark $n\pi^*$ excitations, i.e. they have negligible oscillator strengths when vertically excited.
The order of states at equilibrium geometry is not altered when going to higher levels of theory that include dynamical correlation (see Table~S2 of SI). However, the on-the-fly approach used in this work prohibits the use of higher level methods such as CASPT2 and therefore we employ CASSCF.
The Franck-Condon region --from which excitations take place-- does not only comprise the equilibrium geometry but also slightly distorted geometries. These distortions are due to the different vibrations included within the zero-point energy of the system. Because in cytosine rather small deviations of the equilibrium geometry lead to a different ordering of the state character, the S$_2$ and S$_3$ states can also be bright states and contribute to the absorption spectrum (see Refs.~\cite{Barbatti2010PNAS,Gonzalez-Vazquez2010CPC} and Fig.~S1 of SI). The character of the lowest three excited triplet states at equilibrium is $\pi \pi^*$ for T$_1$ and T$_2$, and $n\pi^*$ for T$_3$. A comprehensive report of vertical excitation energies at different levels of theory can be found in the SI.

In order to obtain a global picture of the relaxation mechanisms of keto-cytosine, we have first used initial conditions spanning the whole first absorbtion band of the UV spectrum, i.e. covering excitation energies from ca 4
to 7 eV. As explained above, this requires launching trajectories from the first three excited states, S$_1$, S$_2$ and S$_3$. Most time-resolved spectroscopic experiments in cytosine~\cite{Kang2002JACS,Canuel2005JCP,Kosma2009JACS,Kotur2011JCP,Ho2011JPCA}
use a pumping wavelength of 267 nm (4.64 eV), just below the center of the first absorption
band located at 260 nm (4.77 eV). In order to narrow the initial conditions to
the energy range corresponding to the experimental one, we have also analyzed
the results (Figure S2 of SI) limited to the bandwidth 4.75$\pm$0.25 eV, just below our
theoretically predicted first absorption band maximum. Also in this energy range, states S$_1$ to S$_3$ are excited. The results from both energy ranges qualitatively agree with each other so that  we will limit
the discussion to the more general broad range.

\begin{figure*}
\centering
\includegraphics[width=0.8\textwidth]{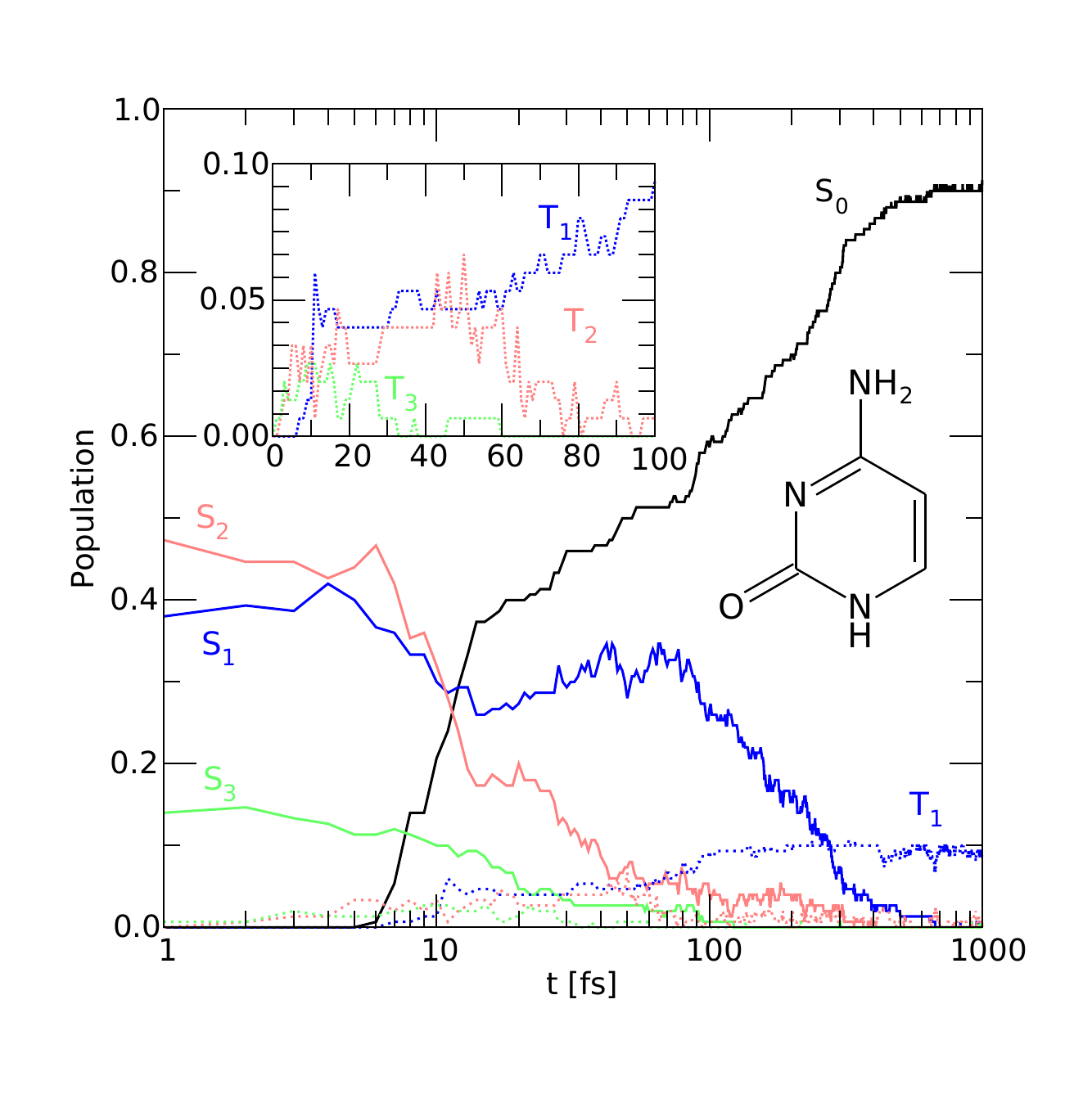}
\caption{Time evolution of the singlet (solid) and triplet (dotted) states
during the first picosecond. The inset zooms the first 100 fs. The S$_0$ ground
state population is in black, S$_1$/T$_1$ states populations are in red,
S$_2$/T$_2$ in blue and S$_3$/T$_3$ in green. }
\label{fig:results:population}
\end{figure*}

\begin{figure}
\centering
 \includegraphics[width=0.5\textwidth]{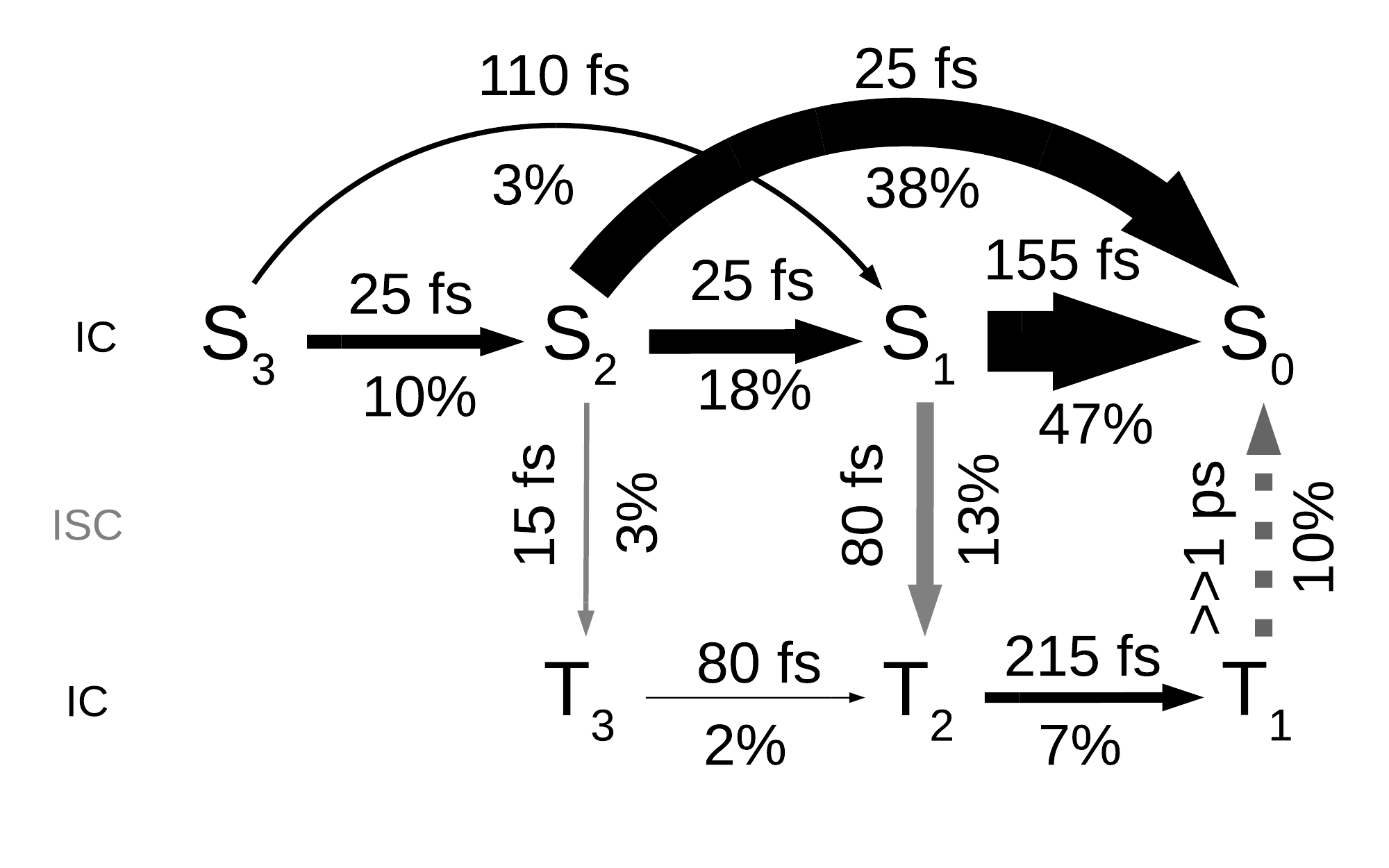}
 \caption{Deactivation pathways of keto-cytosine including internal conversion
(IC, in black) and intersystem crossing (ISC, in grey). The propensity of each
path is sketched by the thickness of the arrows. The dotted line indicates the
deactivation pathway of T$_1$.}
 \label{fig:results:scheme}
\end{figure}

Figure~1 displays the time evolution of all the state populations and Fig.~2 summarizes the most important deactivation paths found in keto-cytosine with \texttt{SHARC}, including decay times and branching efficiencies. One should note that because the calculations are done at levels of theory which do not include sufficient dynamic correlation, the potential energy surface for the dynamics is not accurate enough to derive quantitative conclusions. Since the energy gaps between the states are highly dependent on the level of theory, the branching efficiencies and the decay times should be considered as an informative basis rather than quantitative numbers. We indicate the \textit{total} branching efficiencies over the whole simulation time so that in some cases numbers can be higher than the initial population. Percentages not adding to 100\% are due to minor pathways not indicated. The decay times are obtained fitting the net amount of hops between two particular states (see Fig.~S3 and Table S1 of SI) to an exponential function. The branching efficiencies given in \% are also graphically indicated by arrows of different thickness according to their importance. After photo-excitation, which corresponds to time zero in our simulations,  the population of the $\pi\pi^*$ is distributed as 13\% in S$_3$, 47\% in S$_2$, and
40\% in S$_1$, as dictated by the weight of the oscillator strength of each state. Since the character of a state can adiabatically change during the simulation, hereafter we shall refer to the states by its energetic order rather than by their character.

Analyzing the 13\% population of S$_3$, 10\% relaxes non-adiabatically to the
S$_2$ and from there to the S$_1$ within 25 fs. After ca 155 fs, the system
populates the electronic ground state S$_0$. The remaining 3\% of
the population of S$_3$ deactivates directly to S$_1$ via a threefold degeneracy
S$_3$/S$_2$/S$_1$, as proposed in Ref.~\cite{Kistler2008JCP}. This process is
calculated to be slower than the previous one, with a time constant of 110 fs.
Most of the population in S$_2$ transfers preferably to the lower-lying
electronic states within less than 100 fs. Also in this case the process of IC
is possible via a cascade of subsequent S$_2$/S$_1$ and S$_1$/S$_0$ conical
intersections, or directly via three-state conical intersections
S$_2$/S$_1$/S$_0$, as proposed in Refs.~\cite{Blancafort2004JPCA,Kistler2008JCP}.
Both pathways to the ground state are relevant, in agreement with the time-dependent simulations of
Ref.~\cite{Gonzalez-Vazquez2010CPC}. As deduced from the time constants, the
three-state conical intersection pathway is faster (25 fs) than the two-step
pathway (25 fs / 155 fs). As noticeable from Fig.~1, the first encounter with a
conical intersection takes place within only 10 fs; however, complete
depopulation of the S$_2$ state takes roughly 400 fs.
The S$_1$ state starts with 40\% of population. Loss of population is possible via
IC to the S$_0$ state but due to gains from S$_2$ and S$_3$ the net effect (see Fig.~1) is that ca 30\% of
population remains for ca 100 fs, before decaying exponentially to zero in about 0.5 ps.

Conspicuously, due to near degeneracies between singlet and triplet states, the
trajectories clearly show that besides IC, ISC also takes place during the first
tens of fs. Triplet states are populated strikingly fast (see inset in
Fig.~1). ISC mainly occurs from S$_1$ to T$_2$ but in a lesser extent from S$_2$ to
T$_3$, see Fig.~2 and Figure S3b in SI. In turn, the trajectories in the T$_3$
state relax via IC to the T$_2$ state and those in T$_2$ quickly convert to
T$_1$. A minority of the trajectories deactivate directly from T$_3$ to T$_1$ via a
threefold degeneracy. The simulations show that during the first 20 fs there is a degeneracy of many
states (S$_0$, S$_1$, S$_2$, T$_1$ and T$_2$), allowing for efficient population
transfer. At t=20 fs, T$_1$ and T$_2$ are populated by ca 5\% and T$_3$ by half of it.
The T$_3$ state is depopulated in  less than 100 fs. Most of the population of T$_2$ has
radiationless decayed to T$_1$ within 200 fs but since it is constantly
replenished from S$_1$, a small amount of population persists until 1~ps. The
trajectories show that as a consequence of the ISC process and the subsequent IC
between triplet states, the T$_1$ state has gained significant population (ca
10\%) after 140 fs. This population is stable after the propagation time (1 ps)
and most likely can survive during several ps or even ns~\cite{Ho2011JPCA},
contributing to long-lived transients.
The 10\% of population trapped in the lowest triplet state agrees with the
yield of the dark state (9$\pm$7\%), from which ISC can take place, measured by
Hare et al.~\cite{Hare2007PNAS}.

In general, most of the time decay constants calculated are below 100 fs. Ultrafast time-resolved fs pump-probe experiments~\cite{Peon2001CPL,Pecourt2000JACS,Kang2002JACS,Crespo-Hernandez2004CR,Ullrich2004PCCP,Canuel2005JCP,Pecourt2001JACS,Kosma2009JACS}  for cytosine (and other nucleobases) provide transients with three time scales: one in the fs time regime, $\tau_1 <$  100 fs, and two slower ones, in the ps scale, $\tau_2$ and $\tau_3$ --the latter one even in the ns regime~\cite{Ho2011JPCA}. As recently noted in Refs.~\cite{Kotur2011JCP,Matsika2011FD,KoturIEEE12}, experiments involving ionization integrate over different photoelectron energies or ionic fragments. Interestingly, it has been also demonstrated~\cite{Kotur2011JCP} that different molecular fragments show different time scales (corresponding to very different relaxation pathways to the ground state). Therefore, the three time constants mentioned above correspond most likely to an average over the many deactivation routes taking place in the nucleobases. As such, our calculated time scales for each of the multitude of relaxation pathways can only contribute to the averaged $\tau_1$ decay time measured in Refs.~\cite{Kosma2009JACS,Ho2011JPCA}. Sub-picosecond ISC occurring before vibrational relaxation in the S$_1$ state has been also proposed in DNA base analogues (4-thiothymidine and 6-thioguanosine)~\cite{Reichardt2010JPCL,Reichardt2010CC,Reichardt2011JPCB,Martinez-Fernandez2012CC} lending further support to the participation of triplet states in the relaxation process of cytosine.
The longest  decay time $\tau_3$ can be associated with the ISC from T$_1$ to the ground state S$_0$. Unfortunately, it is difficult to speculate about the origin of the intermediate $\tau_2$ since none of our obtained time scales (recall Figure 2) are in the ps time scale.

Particularly interesting is to note that the obtained ISC mechanism (S$_1 \rightarrow$ T$_2 \rightarrow$ T$_1$)  is different from the one previously proposed in the literature, based solely on quantum chemistry. The calculations of Merch\'{a}n and coworkers~\cite{Merchan2005JACS} propose S$_1 \rightarrow$ T$_1$ transitions. Their calculations find a minimum in the S$_1$ potential from where the S$_1$ and T$_1$ states are almost degenerate along the minimum energy path. A substantial spin-orbit coupling of 20 - 30 cm$^{-1}$ is found along this path~\cite{Merchan2005JACS}.  In their calculations, the S$_1$/T$_1$ crossing corresponds to a transition between the $^1\pi\pi^*$ and
the $^3n\pi^*$ states, in agreement with the El-Sayed selection rules for ISC~\cite{Lower1966CR,Braslavsky2007PAC}.
In contrast, in our calculations, the ISC takes place fundamentally between the $^1n\pi^*$ (S$_1$) and the $^3n\pi^*$  (T$_2$) states, while the $^3\pi\pi^*$ state is the T$_1$. Accordingly, and in agreement with the El-Sayed rules, the spin-orbit coupling between S$_1$ and T$_1$ is larger (ca 15-20 cm$^{-1}$ in average, maximum 40 cm$^{-1}$) than between the S$_1$ and T$_2$ state (ca 5 cm$^{-1}$ in average).
However, the S$_1 \rightarrow$ T$_2$ transition is predominant because the T$_2$ state is separated from the S$_1$ by a very small energy gap ($<$~0.05~eV) for many geometries along the key reaction path. In contrast, the energy gap between the S$_1$ and T$_1$ states is at least 0.1~eV for most of the geometries on the same path and hence hopping is not efficient. Clearly, the close degeneracy between the states of different multiplicity can compensate for very small spin-orbit coupling and drive fast ISC, as recently discussed by Worth and coworkers~\cite{Parker2009CPL,Penfold2010CP} in benzene (see also Ref.~\cite{Marian2012WIRE}).
Thus, the S$_1 \rightarrow$ T$_2$ transition dominates the ISC process. This result illustrates the importance of time-dependent dynamical simulations over quantum chemistry calculations alone. In view of our simulations, the S$_1$  can be considered the precursor state from which the largest ISC takes place before vibrational cooling occurs --as proposed in Ref.~\cite{Hare2007PNAS} or in Refs.~\cite{Hare2006JPCB,Hare2008CP} for uracil and thymine, respectively.

Another important mechanistic conclusion obtained from the present simulations is that ISC coming from higher states also contributes to the deactivation of keto-cytosine. Common belief is that only spin-flips from the low-lying S$_1$ state are relevant because ISC transitions in organic molecules without heavy elements are not considered efficient in the fs timescale. Contrary to that, S$_2 \rightarrow$ T$_3$ transitions are possible in keto-cytosine, leading to population transfer of ca 3\% after few fs. ISC channels from singlet states higher than the S$_1$ have been also suggested in e.g. 4-thiothymidine by Reichardt et al.~\cite{Reichardt2010JPCL} or in uracil and thymine by Etinski et al.~\cite{Etinski2009JPCA}

\begin{figure}
\centering
 \includegraphics[width=0.5\textwidth]{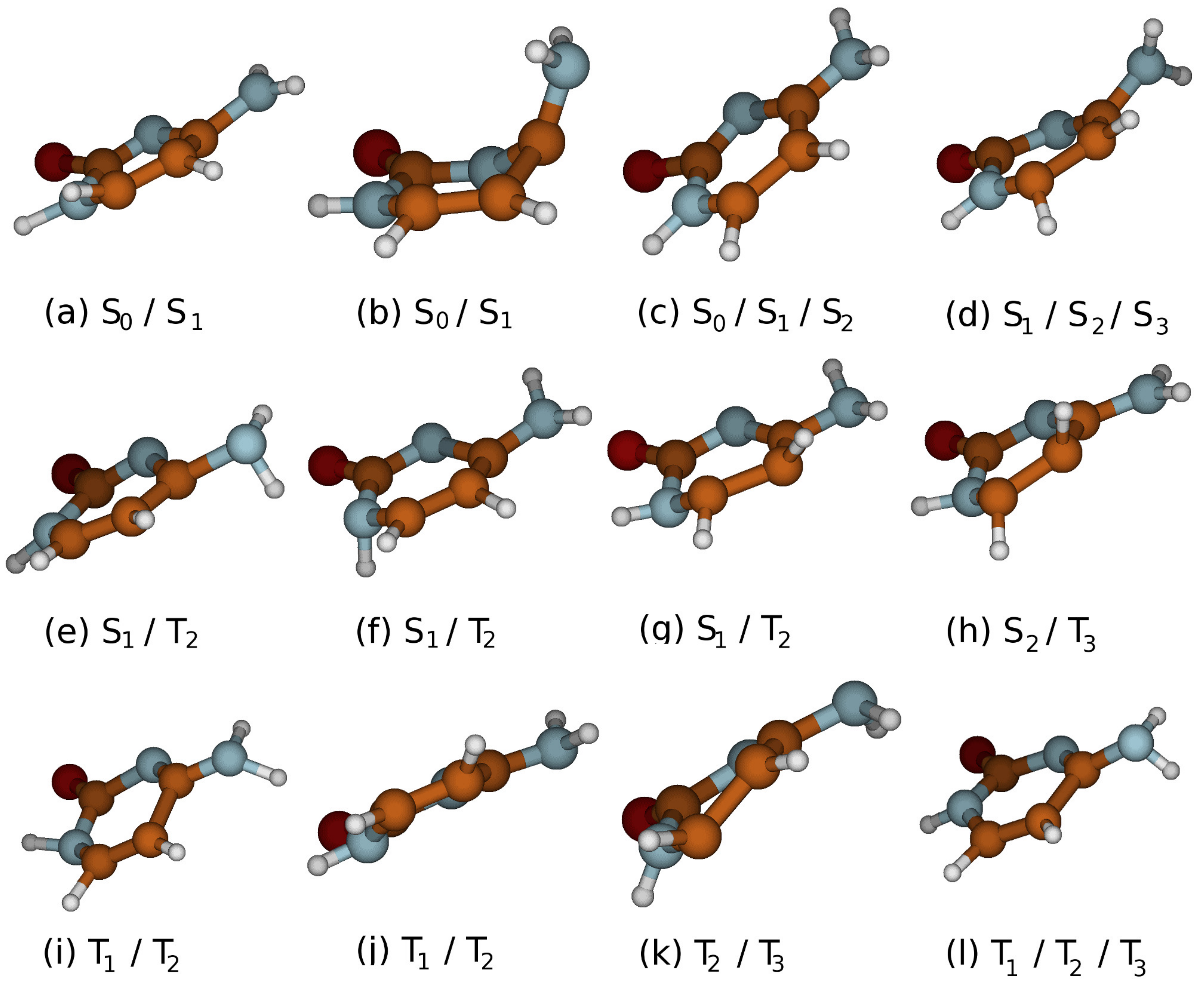}
\caption{Characteristic geometries displaying near degeneracies between two or
three states.}
\label{fig:results:geoms}
\end{figure}

Further insight into the different competing pathways can be obtained by characterizing the molecular geometries that allow for an efficient population transfer.
Figure~3 collects snapshots of archetypal geometries where two or three states are found (near) degenerate (energy less than 0.1 eV). Among the geometries where the S$_0$ and S$_1$ states are close in
energy, two geometries similar to the twist and sofa S$_0$/S$_1$ conical
intersections reported by Kotur et al.~\cite{Kotur2011JCP} could be identified. Figure~3a shows our twist structure, with the carbon atoms twisted by about 70$^{\circ}$ around the C=C double bond. The sofa conical intersection,  Fig.~3b, is characterized by the NH$_2$ nitrogen displaced from the plane, giving the ring structure the appearance of a sofa.
A 3-state near degeneracy between the S$_0$, S$_1$ and S$_2$ states is a frequent event in keto-cytosine, see Fig.~3c. With the C=O bond elongated to 1.42 \AA, the C=C bond between the two CH groups elongated to 1.52 \AA,~and the NH$_2$ group pyramidalized to an angle of 45$^{\circ}$ at the nitrogen, this
geometry resembles the structure first optimized by Blancafort et al.~\cite{Blancafort2004JPCA} and later on by Kistler and coworkers~\cite{Kistler2008JCP}. Interestingly, this structure is also similar to the semi-planar S$_0$/S$_1$ conical intersection published by Barbatti et al.~\cite{Barbatti2011PCCP}.

Figure~3d shows a 3-state degeneracy involving
the S$_1$, S$_2$ and S$_3$ states. Here, the C=C bond is elongated to 1.54 \AA~and the carbon atoms are twisted about 56$^{\circ}$ around the C=C double bond. The dihedral at the NH$_2$ nitrogen is only 22$^{\circ}$ and therefore, the NH$_2$ group is rotated by about 35$^{\circ}$ around the N-C bond rather
than pyramidalized. This geometry is different from the structures proposed by Kistler et al.~\cite{Kistler2008JCP}.
The trajectories show that singlet and triplet states are usually near
degenerate if the NH$_2$ group is pyramidalized and starts rotating around the
C-N bond, as in Fig.~3e, and/or if the hydrogen of the NH group in para position
shows strong out-of-plane oscillations, as in Fig.~3f. Also, when the other
hydrogens of the ring show strong out-of-plane movement, near singlet/triplet degeneracies take
place, see Figs.~3g,h.
To the best of our knowledge no conical intersections  between triplet states
are reported so far --most likely because only T$_1$ was believed to be involved
in ISC.  Figs.~3i,j,k show geometries where two triplet states are  degenerate.
A degeneracy between T$_1$ and T$_2$ (Figs.~3i) occurs when the C-C bond next to
the NH$_2$ group is elongated up to 1.68 \AA. Additionally, the NH$_2$ group is
pyramidalised and the hydrogen of the NH group in para position shows out-of-plane movement.
Other structures like the one given in Fig.~3j show a twist of the C=C double
bond and a slight out-of-plane movement of the whole NH group, leading
to a degeneracy of the T$_1$ and T$_2$ states. The geometries where T$_2$ and
T$_3$ come close in energy show similar features:
pyramidalisation of the NH$_2$ group and the out-of-plane movements of the ring
hydrogens, see Fig.~3k. Additionally, the C=C double bond is twisted and rotated
out-of-plane.
Additionally, triple degeneracies of triplet states have been also identified,
see Fig.~3l, although they are not operative within $1$ ps.
Here, the NH$_2$ group is pyramidalized by 67$^{\circ}$ and the C-C bond next to
the NH$_2$ group is elongated to 1.71\AA. This indicates that IC from T$_3$ is
not only possible via subsequent T$_3$/T$_2$ and T$_2$/T$_1$ conical
intersections, but also directly from T$_3 \rightarrow$ T$_1$.

In summary, our work provides a theoretical rationale for the nuclear factors that make ISC compete with IC in isolated keto-cytosine. As this study demonstrates, the fate of energy relaxation in many complex molecules cannot be revealed by a frozen picture, i.e. quantum chemical calculations at optimized geometries. Instead, a time-dependent picture is necessary, where both IC and ISC compete as dynamical events. ISC is generally quoted in textbooks~\cite{McQuarrie1997} as a slower process in comparison to IC. As a result, many scientists are unaware that ISC can be also an ultrafast process, not only when heavy atoms are present (see e.g.~\cite{CherguiDT12}), but also in organic molecules --as this study evidences.

\begin{acknowledgement}
This work is financed by the Deutsche Forschungsgemeinschaft (DFG) within the
project GO 1059/6-1, by the German Federal Ministry of Education and Research
within the research initiative PhoNa, the Direcci\'on General de Investigaci\'on
of Spain under Project No. CTQ2008-06760, a Juan de la Cierva contract, and the
European COST Action CM0702. We thank T. Weinacht for helpful discussions
regarding cytosine experiments. Generous allocation of computer time in Jena (Computer Center) and Vienna (Vienna Scientific Cluster) is gratefully acknowledged. 
\end{acknowledgement}

\begin{suppinfo}
Computational details, calculated UV absorption spectra as well as additional results.
\end{suppinfo}

\newpage

\textbf{Femtosecond Intersystem Crossing in the DNA Nucleobase Cytosine\\Supporting Information}


\setcounter{figure}{0}
\setcounter{table}{0}
\renewcommand{\thefigure}{S\arabic{figure}}
\renewcommand{\thetable}{S\arabic{table}}

 \subsection{I. Computational details regarding dynamical simulations and initial conditions}

The simulation of the deactivation dynamics of keto-cytosine in gas phase was performed using
the semiclassical \texttt{SHARC} method~\cite{Richter2011JCTC-SI} that calculates non-adiabatic molecular dynamics simulations using Tully's fewest switches criterion \cite{Tully1990JCP-SI}. The nuclei are treated as classical point charges that obey Newton's equations of motion \cite{Newton1687-SI}. To follow the trajectory of the nuclei, the Velocity-Verlet algorithm \cite{Verlet1968PR-SI} was applied with a time step of 0.5 fs. The energy gradients that serve as an input for the Velocity-Verlet
algorithm were analytically calculated in the electronic part of the simulation.
In contrast to the nuclei, the electrons are treated by means of ab initio
quantum mechanics. Here, the electronic wave function is expanded into a linear
combination of basis functions, that represent the contribution of the different
calculated electronic states to the total wave function. The time evolution of
the quantum amplitudes is followed using the 5th order Butcher algorithm
\cite{Butcher1965JACM-SI} with a time step of $5 \cdot 10^{-6}$ fs. The resulting
values were corrected for decoherence effects using the method of Granucci and
Persico with a parameter of $\alpha = 0.1$ hartree \cite{Granucci2007JCP-SI}.

Initial conditions for the trajectories are generated with a Wigner harmonic
distribution of 5000 uncorrelated geometries and velocities.  For this purpose,
the equilibrium geometry was optimized and normal modes were calculated with the
TURBOMOLE package \cite{Ahlrichs1989CPL-SI, Treutler1995JCP-SI}, the B3LYP hybrid
functional~\cite{Becke1993JCP-SI,Lee1988PRBa-SI} and the TZVP basis set
\cite{Schafer1994JCP-SI}.
For each of the initial conditions, a single-point calculation employing the
Complete Active Space Self-Consistent-Field
approach~\cite{Werner1985JCP-SI,Knowles1985CPL-SI} averaged over four singlet states
with the MOLPRO\cite{Werner-SI} package and the 6-31G* basis set was done (see
Section III for further electronic structure details). All 5000 initial
conditions and their calculated oscillator strengths and vertical excitations
were used to create a UV absorption spectrum as described in
Ref.~\cite{Barbatti2010PCCP-SI}. The spectrum, see Fig. S1, is
in good agreement with that previously published in
Ref.~\cite{Gonzalez-Vazquez2010CPC-SI}.

\begin{figure*}[ht]
\centering
 \includegraphics[width=0.5\textwidth]{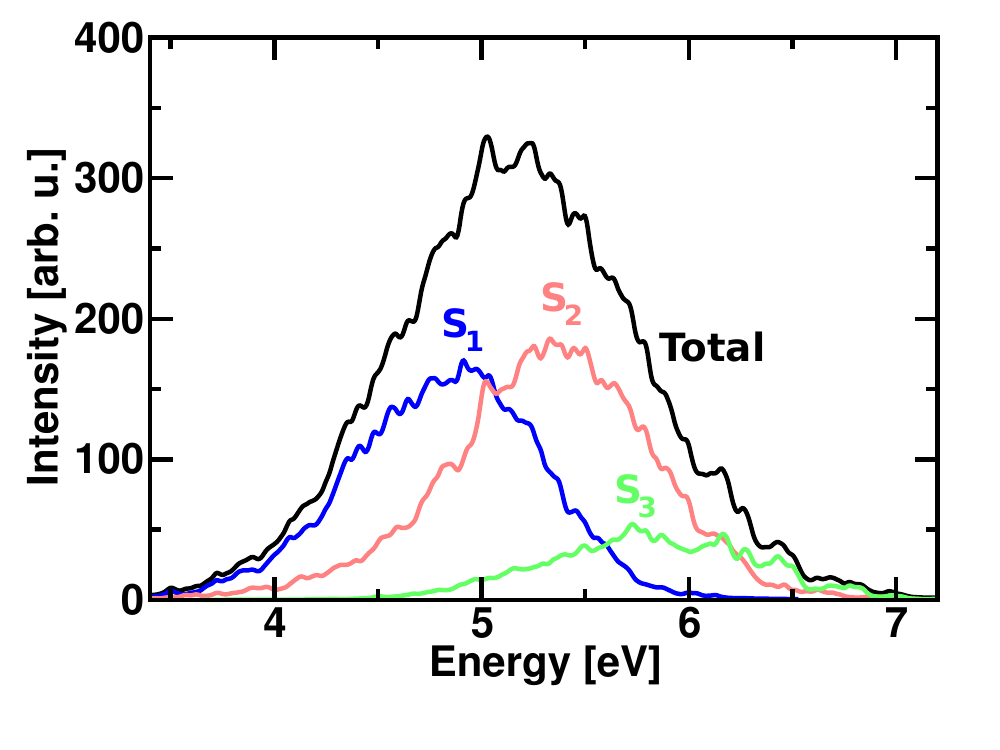}
 \caption{UV absorption spectrum of keto-cytosine.}
 \label{fig:si:spectrum}
\end{figure*}

Using Newton-X \cite{Barbatti2007JPPA-SI, NX-program07-SI} it is possible to estimate
the instantaneous probability to excite keto-cytosine at every geometry to a
particular excited state. Applying these probabilities to the calculated 5000
initial conditions, we obtain 591 trajectories starting from the S$_1$, 685
starting from the S$_2$ and 192 starting from the S$_3$ state. Since the
molecular dynamics simulations converge with notably less than 1468
trajectories, only 150 trajectories have been propagated. This ensemble
consisted of 60, 70 and 20 trajectories, starting from the S$_1$, S$_2$ and
S$_3$ respectively. The results were statistically evaluated for all
trajectories and checked for convergence.
Each trajectory was propagated during 1 ps using time steps of 0.5 fs.

Population transfer of the semiclassical trajectories between the different
electronic states was calculated according to the Tully's fewest switches
criterion~\cite{Tully1990JCP-SI} in the adiabatic representation, including
non-adiabatic couplings as well as spin-orbit couplings. 
In order to avoid unphysical hops and for computational reasons, the
hopping probabilities were only calculated for states that, after a hop, would
lead to an increase of potential energy of 1 eV or less. After a hop, the
kinetic energy was adjusted in order to conserve the total energy of the system.
Therefore, the velocities (i.e. the kinetic energies) of the atoms were scaled
along their current direction in order to keep the total energy constant.

If a trajectory populated the S$_0$ state for at least 20 fs with a coefficient
greater than 0.99, it was stopped automatically and it was assumed that the
population in the S$_0$  remains constant for the rest of the simulation.

\subsection{II. Deactivation pathways with an excitation bandwidth of 4.75$\pm$0.25 eV}
Figure~S2 displays the time evolution of all the state populations when limiting the excitation bandwidth to 4.75$\pm$0.25 eV, in accord to the energy range corresponding to the experiments, see e.g. Ref. ~\cite{Kosma2009JACS-SI}. In this energy range only 42 trajectories are considered. As it can be seen the results are qualitatively the same as those shown in Fig.~1 of the manuscript using 150 trajectories in the broader energy range of 4 to 7 eV. After 1000 fs the populations are converged to the same ratio, about 90\% in S$_0$ and 10\% in T$_1$. Moreover, the S$_0$ gains about 40\% population within 10-15 fs. Obviously, the initial distribution of the trajectories is changed with respect to the full picture with 150 trajectories: The S$_1$ is now the strongest populated state in the beginning, while S$_2$ and S$_3$ states are less populated and therefore hit zero population faster. However, the early population of the triplet states and the deactivation channels, and therefore, the main conclusions discussed in the main text are the same as in the broader range. The reason why the two sets of energy ranges give very similar dynamical results is based on the fact that the S$_1$ state provides the most important channel for the ultrafast ISC observed in our simulations. The branching between the two main deactivation pathways from S$_1$ (i.e. ISC towards T$_2$ and following IC towards T$_1$ or IC from S$_1$ to S$_0$) is similar in the complete range of initial conditions. Then, because S$_1$ and T$_2$ are nearly degenerate for many geometries, the smaller excess energy of the trajectories with limited excitation bandwidth does not influence the efficiency of the ISC channel. Thus, the early population of the triplet states and the deactivation channels discussed in the main text are almost independent of the excitation energy range.

The deactivation pathways visualized in Fig. 2 are based on the number of
surface hops that occurred between the different pairs of states in the
molecular dynamics simulations. Since jumps can take place in two directions,
the net amount of hops for a pair of states $i$ and $j$ can be calculated by
subtracting the number of hops from state $i$ to $j$ from the number of hops
from $j$ to $i$. The temporal evolution of this quantity is shown in Fig.~S3.
Panel a) depicts the pathways for internal conversion (IC) between the singlet
states, panel b) the intersystem crossing (ISC) pathways between singlet and
triplet states and panel c) the IC between the triplets. The curves can be
fitted with exponential functions according to  $N_{i} (1-e^{-t/\tau_{i}})$ in
order to get time constants $\tau_{i}$ for the respective pathways. These time
constants are collected in Table~S1 and are consistent with all the available data.
However, due to the sensitivity of the time constants with the energy and the
lack of dynamical correlation in the electronic structure calculations,
they should be regarded as qualitative.

\begin{figure*}[ht]
\centering
  \includegraphics[width=0.7\textwidth]{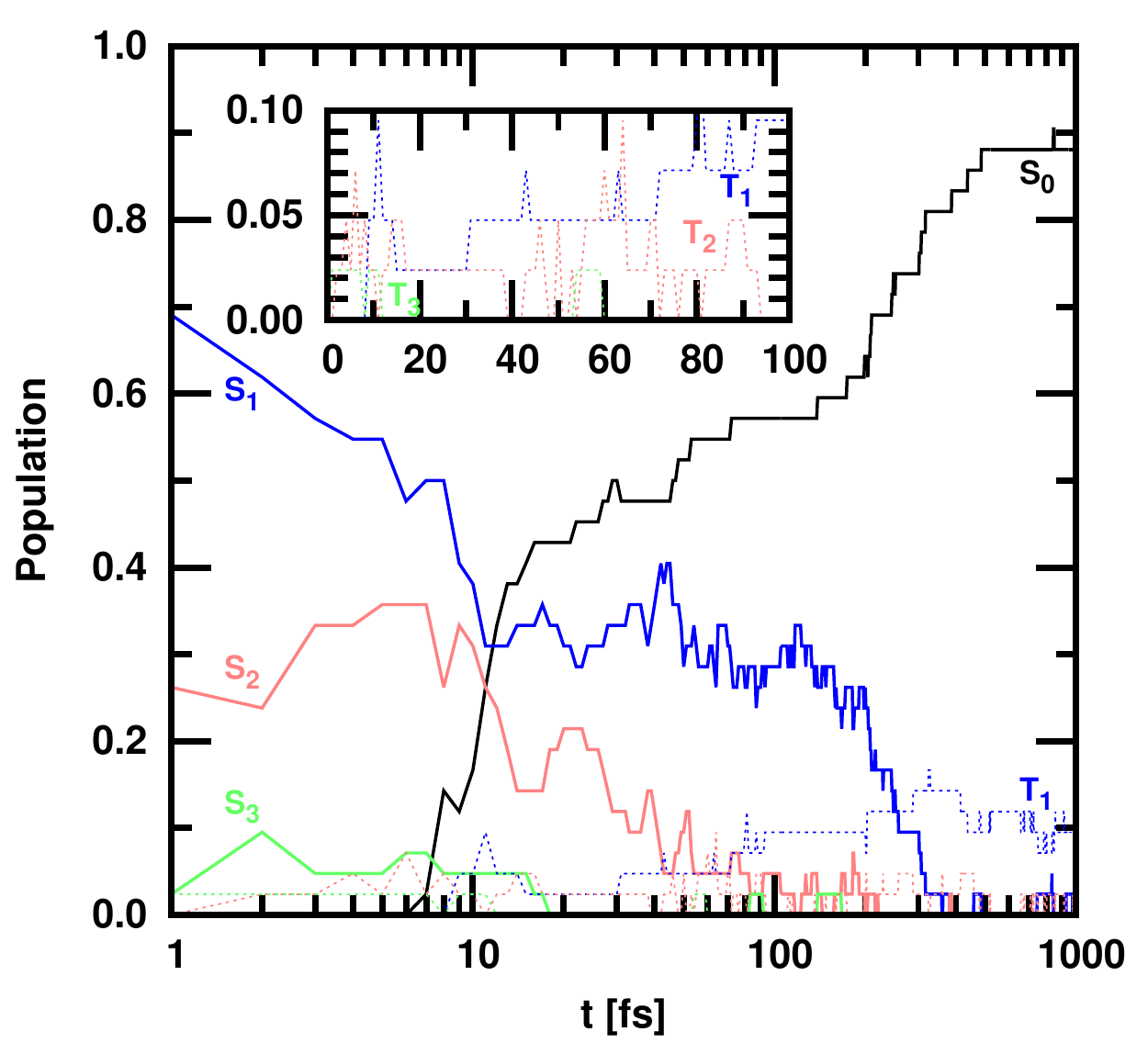}
 \caption{Time evolution of the singlet (solid) and triplet (dotted) states
during the first picosecond. The inset zooms the first 100 fs. The S$_0$ ground
state population is in black, S$_1$/T$_1$ states populations are in red,
S$_2$/T$_2$ in blue and S$_3$/T$_3$ in green. Excitation bandwidth is limited to 4.75$\pm$ 0.25 eV.}
 \label{fig:si:limitedbandwidth}
\end{figure*}

\begin{figure*}[ht]
\centering
  \includegraphics[width=0.95\textwidth]{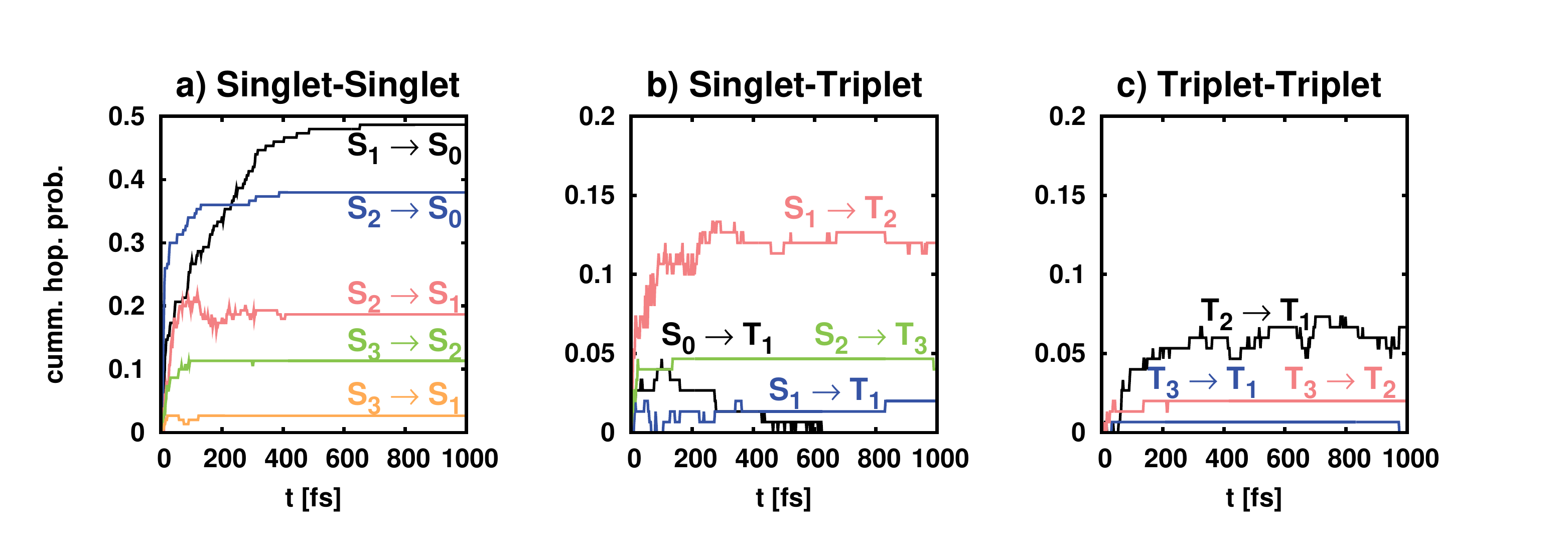}
 \caption{Net hops accumulated in time.}
 \label{fig:si:cummhops}
\end{figure*}

\begin{table*}[t]
\centering
\caption{Time constants fitted to accumulated hopping probability curves (see Fig.
S2)}
\begin{tabular}{lll}
Pathway                   & time constant        & Error [\%]\\
\hline\hline
S$_1$ $\rightarrow$ S$_0$ & $\tau_{1}=$ 155 fs   & 1 \\
S$_2$ $\rightarrow$ S$_0$ & $\tau_{2}=$ 25 fs    & 2 \\
S$_2$ $\rightarrow$ S$_1$ & $\tau_{3}=$ 25 fs    & 3 \\
S$_3$ $\rightarrow$ S$_1$ & $\tau_{4}=$ 109 fs   & 3 \\
S$_3$ $\rightarrow$ S$_2$ & $\tau_{5}=$ 22 fs    & 2 \\
\hline
S$_1$ $\rightarrow$ T$_1$ & $\tau_{6}=$ 1000 fs  & $>$100 \\
S$_1$ $\rightarrow$ T$_2$ & $\tau_{7}=$ 78 fs    & 2 \\
S$_2$ $\rightarrow$ T$_3$ & $\tau_{8}=$ 14 fs    & 4 \\
\hline
T$_2$ $\rightarrow$ T$_1$ & $\tau_{4}=$ 216 fs   & 3 \\
T$_3$ $\rightarrow$ T$_1$ & $\tau_{5}=$ 33 fs    & 2 \\
T$_3$ $\rightarrow$ T$_2$ & $\tau_{5}=$ 82 fs    & 2 \\
\end{tabular}
\label{tab:si:timeconstants}
\end{table*}

\clearpage

\subsection{III. Ab initio quantum chemical calculations}
At each time step of the propagation, the electronic states of keto-cytosine and
corresponding electronic properties have been calculated using the state-average
SA7-CAS(12,9)/6-31G* level of theory using 12 electrons distributed
in 9 orbitals, as implemented in the MOLPRO software package~\cite{Werner-SI}. As
in Ref.~\cite{Gonzalez-Vazquez2010CPC-SI}, the active space consists of 4 $\pi$, 3
$\pi^*$ and 2 $n$ orbitals, see Fig.~S4. Seven states
(four singlets and three triplets) are averaged with equal weigths. Relativistic
corrections were taken into account by the use of the 2nd order
Douglas-Kroll-Hess Hamiltonian \cite{Douglas1974AP-SI, Hess1986PRA-SI}.

\begin{figure*}[htb]
\centering
 \includegraphics[width=0.35\textwidth]{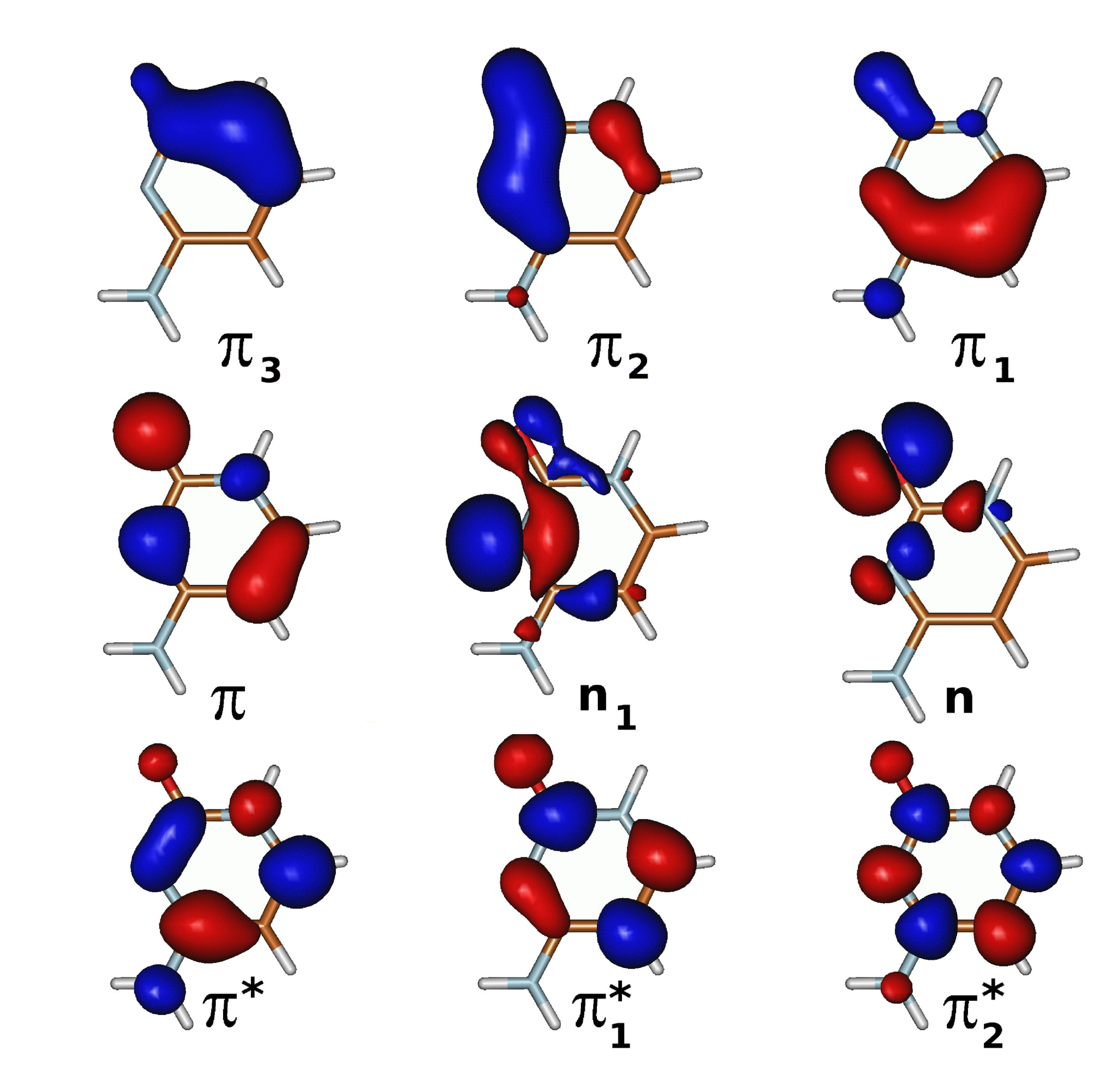}
 \caption{Orbitals included in the SA7-CASSCF(12,9)/6-31G* calculations.}
 \label{fig:si:active-space}
\end{figure*}

Table~S2 shows the vertical excitation energies of the
lowest-lying three singlet and triplet excited states at the equilibrium
geometry obtained at SA7-CASSCF(12,9)/6-31G*//B3LYP/TZVP level of theory as well
as other levels of theory for comparison, obtained using MOLCAS
7.6~\cite{molcas1-SI,molcas2-SI,molcas3-SI}. As it can be seen, the energies are rather
sensitive to the introduction of dynamical correlation and also to the optimized
geometry employed. The  values obtained with SA7-CASSCF(12,9)/6-31G*//B3LYP/TZVP
are similar to those obtained when calculating the singlets and triplet states
separately (SA4+SA3). The energy of the lowest singlet state S$_1$, which
determines the maximum of the first UV absorption band, is decreased when perturbation theory is included (MS-CASPT2/SA7-CASSCF(12,9)/6-31G*//B3LYP/TZVP).

At MP2 level of theory the ground state equilibrium structure gets slightly
pyramidalized at the NH$_2$ group while B3LYP keeps the molecule
completely planar. This geometrical difference is not relevant in the ab initio
MD simulations, since an ensemble of geometries sampled over all the degrees of
freedom according to the ZPE is considered. However, as it can be seen in
Table~S2, the out of plane deformation is important to obtain an
energy closer to the experimental S$_1$-origin in the gas phase~\cite{Nir2002CPL-SI,Nir2002EPJD-SI}
located at ca. 4.7 eV. The MS-CASPT2 value for the S$_1$ using
SA4+SA3-CAS(12,9)/6-31G*//MP2/6-31G* level of theory is predicted at 4.8 eV,
very close to the experimental one. A similar value is also obtained using
DFT-MRCI calculations~\cite{Tomic2005JPCA-SI}.

Since MS-CASPT2 trajectories are computationally not affordable and electronic couplings are not available at DFT-MRCI level of theory, our calculations were done using SA7-CAS(12,9)/6-31G*//B3LYP/TZVP. The obtained
vertical excitation energies are in agreement with those previously reported
also at the CAS/6-31G** level of theory by Merch\'{a}n and coworkers~\cite{Merchan2005JACS-SI}.

\begin{table*}[ht]\footnotesize
\centering
\caption{Vertical excitation energies of keto-cytosine at the
equilibrium geometry computed at different levels of theory, as specified.
Oscillator strengths are given in parentheses.}
\begin{tabular}{c||ccc|ccc}
{\it ab initio} methodolody
& S$_1$ & S$_2$ & S$_3$ & T$_1$ & T$_2$ & T$_3$ \\
\hline\hline
SA7-CAS(12,9)/6-31G*//B3LYP/TZVP &
5.37 (0.081) & 5.38 (0.002) & 5.79 (0.002) &
3.81 & 5.14 & 5.17 \\
SA4+SA3-CAS(12,9)/6-31G*//B3LYP/ZVP &
5.26 (0.083) & 5.41 (0.001) & 5.72 (0.004) &
3.80 & 4.96 & 5.02 \\
MS-CASPT2 using previous one &
5.00 (0.073) & 5.37 (0.002) & 5.81 (0.002) &
4.15 & 5.17 & 5.29 \\
\hline
SA7-CAS(12,9)/6-31G*//MP2/6-31G* &
5.08 (0.074) & 5.09 (0.000) & 5.46 (0.005) &
3.54 & 4.91 & 4.93 \\
SA4+SA3-CAS(12,9)/6-31G*//MP2/6-31G* &
4.98 (0.080) & 5.10 (0.001) & 5.45 (0.004) &
3.48 & 4.74 & 4.77 \\
MS-CASPT2 using previous one &
4.80 (0.072) & 5.15 (0.002) & 5.61 (0.002) &
3.96 & 4.90 & 5.17 \\
\hline\hline
CAS(12,9)/6-31G** Ref~\cite{Merchan2005JACS-SI} &
5.32 & 5.34 & 5.67 &
3.72 & 5.01 & 5.19 \\
MS-CASPT2 Ref~\cite{Merchan2005JACS-SI} &
4.53 (0.065) & 5.04 (0.001) & 5.11 (0.003) &
3.65 & 4.68 & 4.77\\
\hline\hline
DFT-MRCI//B3LYP/TZVP Ref~\cite{Tomic2005JPCA-SI} &
4.83 (0.080) & 5.02 (0.002) & 5.50 (0.001) &
& & \\
\hline
\end{tabular}
\label{tab:si:vertical}
\end{table*}

\end{document}